\documentclass[12pt,epsfig]{article}
\usepackage{graphicx}
\usepackage{dcolumn}
\usepackage{bm}
\usepackage{epsfig}
\usepackage{color}
\usepackage[a4paper]{geometry}
\newcommand{\beq}{\begin{equation}}
\newcommand{\eeq}{\end{equation}}
\newcommand{\beqarray}{\begin{eqnarray}}
\newcommand{\eeqarray}{\end{eqnarray}}
\begin{document}
\begin{center}
{Cosmic Ray Antiprotons from Nearby Cosmic Accelerators} 
\\
\medskip
{Jagdish C. Joshi and Nayantara Gupta}
 
{Astronomy and Astrophysics Group, Raman Research Institute, Sadashivanagar, Bangalore 560080, INDIA}
\footnote{jagdish@rri.res.in}
\end{center}
\begin{abstract}
The antiproton flux measured by PAMELA experiment might have originated from Galactic sources of cosmic rays. These antiprotons are expected to be produced in 
the interactions of cosmic ray protons and nuclei with cold protons. 
Gamma rays are also produced in similar interactions inside some of the cosmic acceleratos. We consider a few nearby supernova remnants observed by Fermi LAT. 
Many of them are associated with molecular clouds.
Gamma rays have been detected from these sources which most likely originate in decay of neutral pions produced in hadronic interactions. The observed gamma ray fluxes from these SNRs are used to find out their contributions to the observed diffuse cosmic ray antiproton flux near the earth.
\end{abstract}

Keywords:cosmic rays, antiprotons, gamma rays\\

\section{Introduction}

Antiprotons are secondary particles produced in the interactions of primary cosmic rays with cold protons. The primary cosmic ray composition in the energy range of 10-1000GeV is dominated by protons \cite{pamela1,pamela2}.
PAMELA \cite{pamela3} experiment has measured the flux of cosmic ray antiprotons  in the energy range of 60 MeV to 180 GeV. These particles could be produced
in the interactions of cosmic ray protons with the cold protons inside the cosmic accelerators \cite{blasi1,bereh,fujita} and also in the interstellar medium during the propagation of cosmic rays \cite{don,shibata1,shibata2}.
 Although antiprotons are much less in number compared to protons they could be the messengers of unknown interactions of particles leading to the production of
antiparticles. A better understanding of the antiproton flux would be useful to explore such possibilities.
\par
In this paper we discuss that the gamma ray emission from the hadronic accelerators close to us can be used to find out the contributions of these sources to the diffuse cosmic ray antiproton flux measured near the earth.
We have considered some nearby SNRs observed in gamma rays by Fermi LAT, many of them are associated with molecular clouds \cite{tanaka,abdo1,ack,abdo2,abdo3,cas,gio}.
The gamma ray fluxes observed from these sources are most likely produced in hadronic interactions of cosmic ray protons ($p\,p\rightarrow \pi^0 \rightarrow \gamma \, \gamma$).

\par
Antiprotons are also produced in hadronic interactions of cosmic ray protons ($p\,p\rightarrow p\,\bar p\,p\,p $). These antiprotons may annihilate and interact with the cold
protons inside the sources and some of them will escape to the interstellar medium. Due to the low density of hydrogen or cold protons in the interstellar medium diffusion loss of cosmic ray antiprotons is more important than the loss due to interactions.
\par

We have used the cross-sections of production of antiprotons from \cite{shibata1}. The simple formalism discussed in this paper can be applied to any hadronic cosmic ray source from which the gamma ray flux has been measured.
Ahlers et al. \cite{ahlers} considered the production of electrons and positrons in SNRs in hadronic interactions. In case of hadronic interactions the observed gamma ray flux produced in neutral pion decay can be used to normalize the primary cosmic ray flux and to obtain the fluxes of other secondary particles.

\section{Antiprotons and Gamma Rays from Nearby SNRs}

The cosmic ray density of the protons inside the sources is expressed by a power law with spectral index $\alpha$, $\frac{dQ_p(E_p)}{dE_pdVdt}=C_p E_p^{-\alpha}$. These cosmic 
rays interact with the ambient cold protons in the molecular clouds producing charged and neutral pions. The charged pions decay to neutrinos/antineutrinos and electrons/positrons.
The neutral pions decay to gamma rays. Antiprotons are also produced in the interactions of cosmic ray protons with the cold protons but with a different cross-section of interaction. 
The antiproton flux injected from cosmic ray interactions is 
\beq
\frac{dQ^{inj}_{\bar p}(E)}{dEdVdt}=2 \,t_{esc}^{inner}\, \rho_s \, c \,\int_{E}^{\infty} \sigma_{pp\rightarrow \bar p\,X} (E,E_p) \frac{dQ_p(E_p)}{dE_pdVdt}\frac{dE_p}
{E_p}.
\label{antip_flux1}
\eeq
The factor of 2 accounts for the contribution from antineutrons equally produced in $p\,p$ interactions.
The speed of the relativistic cosmic rays is close to the speed of light $c$.
The average time of escape for the cosmic rays from the SNR-molecular cloud region is $t_{esc}^{inner}$ sec, average number density of cold protons in the inetracting
medium is $\rho_s$cm$^{-3}$.
\par 
 Following the simplifications given in \cite{shibata1} for high energy antiprotons Eqn.(\ref{antip_flux1}) reduces to
\beq
\frac{dQ^{inj}_{\bar p}(E)}{dEdVdt}=2 t_{esc}^{inner}\rho_s \, c \, \bar\sigma_{\alpha}(E) \, C_p \,E^{-\alpha}
\label{simple_shibata}
\eeq
 The values of $\bar\sigma_{\alpha}(E)$ are given in Fig.4. of \cite{shibata1} for a power law spectrum of cosmic rays with spectral index $\alpha=$2.6 to 2.8.
We have used the cross-sections corresponding to spectral index 2.6 in our calculations.
The diffusion of cosmic rays \cite{ginz} for a spherically symmetric geometry can be used to obtain the propagated or observed cosmic ray antiproton flux. 

The diffusion coefficient $D(E)=D_0 \Big(E/E_0\Big)^{\delta}$ is assumed to be only energy dependent with $\delta=0.33$ above the break at $E_0=4 GeV$ and $D(E)=D_0\sim 10^{28}\frac{cm^2}{sec}$
below $E_0=4GeV$. The cosmic ray antiprotons interact with the cold protons inside the source and in the interstellar medium. The annihilation cross-section of antiprotons with protons 
steeply falls off above 1 GeV while the $\bar p p$ inelastic cross-section rises at the same energy. The total $\bar p p$ interaction cross-section is nearly constant at high energy.
 The formalism discussed in this work uses spherical symmetry around the source. If the source is located close to us and it is well inside the galactic halo the effect of the boundaries of galactic halo may be neglected. Near the eath the observed cosmic ray flux is in steady state.
The source is emitting continuosly. The spherical volume containging the cosmic rays is expanding as a result cosmic ray density is falling inside this volume. At the same time new cosmic rays are arriving from the source. If the loss and gain are compensated then a steady state cosmic ray flux is maintained at a distance R from the source.
 
Following eqn.(9) of \cite{aha1} the maximum antiproton flux at a distance $R$ 
is
\beq
J^{ob}_{\bar p}(E)=c \frac{dQ^{inj}_{\bar p}(E)}{dEdVdt} \frac{V_{source}}{(4\pi)^2\,D(E)\, R} 
\label{antip_flux}
\eeq 
This flux is isotropic with dimension 
GeV$^{-1}$ cm$^{-2}$ sec$^{-1}$ sr$^{-1}$. $V_{source}$ is the volume of the source. D(E) is the energy dependent diffusion coefficient and R is the distance to the source. 
The gamma ray flux density produced inside a source in cosmic ray interactions is
\beq
\frac{dQ^{inj}_{\gamma}(E)}{dEdVdt}=2 t_{esc}^{inner} \frac{ \rho_s \, c}{K_{\pi}} \,
\int_{E_{min}} ^{\infty}
 \sigma_{pp\rightarrow \gamma \gamma}\Big(m_p+\frac{E_{\pi}}{K_{\pi}}\Big) \frac{dQ_{p}(m_p+E_{\pi}/K_{\pi})}{dE_{p}dVdt} \frac{dE_{\pi}}{\sqrt{E_{\pi}^2-m_{\pi}^2}}
\label{gamma_flux1}
\eeq
from eqn.(78) of \cite{kelner}, where $E_p=m_p+E_{\pi}/K_{\pi}$.
The minimum energy of the pions $E_{min}=E+m_{\pi}^2/4E$ and $K_{\pi}=0.17$ gives the fraction of the proton's energy going to the pion.
The gamma ray flux received on earth is $J^{ob}_{\gamma}(E)$
(GeV$^{-1}$ cm$^{-2}$ sec$^{-1}$).
\beq
J^{ob}_{\gamma}(E)=\frac{dQ^{inj}_{\gamma}(E)}{dEdVdt}\frac{V_{source}}{4\pi 
R^2}
\label{gamma_flux2}
\eeq
The cross-section of $p\,p$ interactions for the production of neutral pions is
\beq
\sigma_{pp\rightarrow \gamma \, \gamma}(E_p)=34.3+1.88 \ln(E_p/1TeV)+0.25\ln(E_p/1TeV)^2 \Big[1-\Big(\frac{E_{th}}{E_p}\Big)^4\Big]^2 mb.
\label{cross_sec}
\eeq
where $E_{th}=m_p+2m_{\pi}+m_{\pi}^2/2m_p=1.22 GeV$. In our calculation this cross section is constant $\sim$30 mb in the energy range of 10 GeV-1000 GeV.
Using eqn.(\ref{antip_flux}) and eqn.(\ref{gamma_flux2}) the ratio of the observed fluxes of antiprotons and gamma rays is
\beq
Ratio(E)=
\frac{J^{ob}_{\bar p}(E)}{J^{ob}_{\gamma}(E)}= 
\frac{c \, R \,\frac{dQ^{inj}_{\bar p}(E)}{dEdVdt}}{4\pi \, D(E)\,\frac{dQ^{inj}_{\gamma}(E)}{dEdVdt}}.
\label{obs_flux_ratio}
\eeq

 With eqn.(\ref{simple_shibata}) to eqn.(\ref{cross_sec}) we simplify 
eqn.(\ref{obs_flux_ratio}) to 
\beq
Ratio(E)=\frac{J^{ob}_{\bar p}(E)}{J^{ob}_{\gamma}(E)}= 
\frac{c\, R\, K_{\pi}\, \alpha}{4\,\pi\,D(E)}\times \frac{\bar{\sigma}_{\alpha}(E)}{\sigma_{pp\rightarrow \gamma \gamma}}
\label{ratio}
\eeq
The observed gamma ray fluxes from the individual sources considered in this work are multiplied with the ratio given in eqn.(\ref{ratio}) to obtain the cosmic ray antiproton fluxes from each of them. Our calculated antiproton fluxes are shown in Fig.1. and compared with the antiproton flux observed by PAMELA \cite{pamela3}. In our case $\alpha$ is varying in the range of 1.85 to 3.02 as shown in Table-I. 
 The proton fluxes calculated for the SNRs associated with molecular clouds are shown in Fig.2. assuming their escape time from the molecular clouds $t_{esc}^{inner}=1000$ years. The density of the molecular clouds is assumed to be $\rho_s=100$ cm$^{-3}$ which corresponds to $p\,p$ interaction time $t_{pp}=6\times 10^5$ years  \cite{gabici}. In this case the escape time is much smaller than the $pp$ interaction time so the proton spectra are not attenuated significantly.
\begin{center}
TABLE-I
\vskip 0.5cm
Nearby SNRs considered in the present work 
\vskip 0.25cm
\begin{tabular}{|l|l|l|l|l|}
\hline
SNR & R & Energy range & Gamma ray flux & Ref\\
 & kpc& GeV & $GeV^{-1} cm^{-2} sec^{-1}$& \\
\hline
 Vela Jr & 0.75 & 1-300  &  8.65 $\times 10^{-9} E^{-1.85}$ & \cite{tanaka}\\
 IC443 & 1.5 & 0.2-3.25  &  6.12 $\times 10^{-8} E^{-1.93}$ & \cite{abdo1}\\
 &  & Above 3.25 &  1.29 $\times 10^{-7} E^{-2.56}$ & \cite{ack} \\
 W28& 2 & 0.4-1 & 4.66 $\times 10^{-8} E^{-2.09}$& \cite{abdo2} \\
 &  & Above 1 & 4.66 $\times 10^{-8} E^{-2.74}$ &  \\
  W44 & 3 & 0.1-1.9  & 1.15 $\times 10^{-7} E^{-2.06}$ & \cite{abdo3}\\
  &  & Above 1.9  & 2.13 $\times 10^{-7} E^{-3.02}$ & \cite{ack}  \\
 W30 & 4.5 & 0.1-100 & 2.16 $\times 10^{-8} E^{-2.4}$ & \cite{cas}\\
 Tycho & 1.7-5 & 0.4-100 & 1.38 $\times 10^{-9} E^{-2.3}$ & \cite{gio} \\
\hline
\end{tabular}
\end{center}
The parameters used in this table can also be accessed from \cite{para}.

\vspace{0.9cm}

\begin{figure*}[t]
\centerline{\includegraphics[angle=-90,width=3.5in]{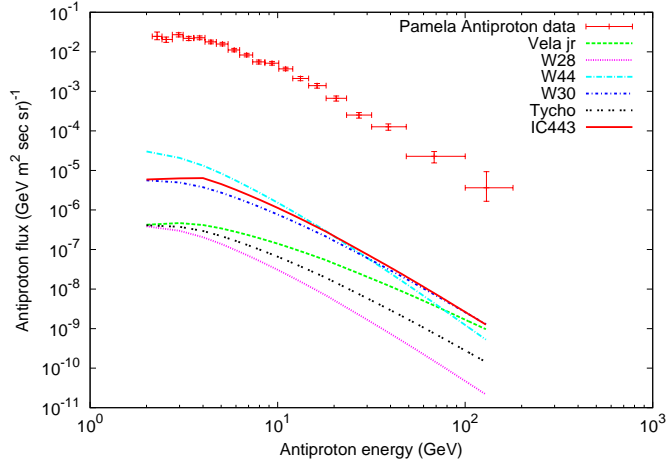}}
\caption{Antiproton fluxes from hadronic cosmic accelerators close to us compared with the total flux observed by PAMELA \cite{pamela3}}
\label{flux_ratio}
\end{figure*} 

\begin{figure*}[t]
\centerline{\includegraphics[angle=-90,width=3.5in]{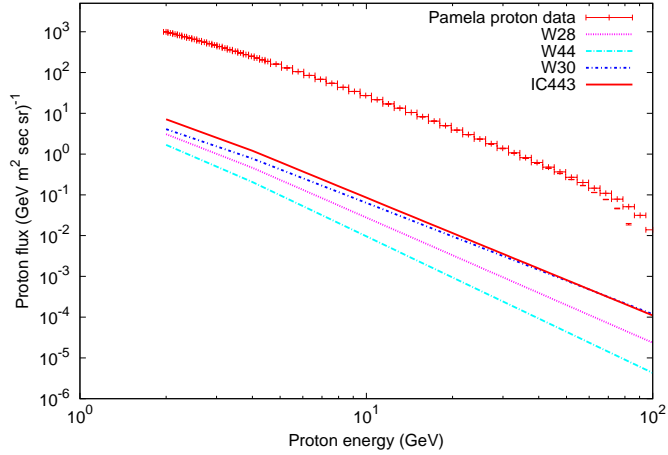}}
\caption{Proton fluxes from the molecular clouds associated with SNRs compared with the total flux observed by PAMELA \cite{pamela1,pamela2} for the following values of parameters $t_{esc}^{inner}=1000$ years, $\rho_s=100$ cm$^{-3}$.}
\label{flux_ratio}
\end{figure*} 

\section{Summary and Conclusion}
A large number of gamma ray point sources have been detected by Fermi LAT and other gamma ray detectors. In some of these sources hadronic interaction of cosmic rays is the underlying mechanism of gamma ray production. We have discussed a simple formalism to find the antiproton fluxes produced inside SNRs and molecular clouds in cosmic ray interactions ($p\,p$) using the gamma ray fluxes from these sources produced in hadronic interactions ($p\,p$) through the decay of neutral pions. The cosmic ray antiproton fluxes originating in $p\,p$ interactions from nearby cosmic accelerators are found to be much less compared to the flux observed by PAMELA \cite{pamela3}.    
 We have assumed energy independent escape of the cosmic ray antiprotons from the sources. Our calculated spectra have energy dependence qualititavely similar to the observed antiproton and proton spectra shown in Fig.1. and Fig.2.. Thus this assumption does not contradict the observational results. We have shown the spectra above 2 GeV as at lower energy the effect of solar modulation becomes important.


\begin{thebibliography}{99}

\bibitem{pamela1} O. Adriani et al. (PAMELA Collaboration), Science {332}, 69 (2011).
\bibitem{pamela2} O. Adriani et al. (PAMELA Collaboration), ApJ {\bf 765}, 91 (2013)

\bibitem{pamela3} O. Adriani et al. (PAMELA Collaboration), PRL {\bf 105}, 121101 (2010).
\bibitem{blasi1} P. Blasi, P. D. Serpico, PRL {\bf 103}, 081103 (2009).
\bibitem{bereh} E. G. Berezhko, L. T. Ksenofontov, arxiv:1405.5281.
\bibitem{fujita} Y. Fujita et al., Phys. Rev. D {\bf 80}, 063003 (2009).
\bibitem{don} F. Donato et al., ApJ {\bf 563}, 172 (2001).
\bibitem{shibata1} T. Shibata, Y. Futo and S. Sekiguchi, ApJ {\bf 678}, 907 (2008).
\bibitem{shibata2} T. Shibata and Y. Futo, Aph {\bf 29}, 907 (2008).

\bibitem{tanaka} T. Tanaka et al. ApJ {\bf{740}} L51 (2011).
\bibitem{abdo1} Abdo et al. ApJ {\bf{712}}, 459 (2010).
\bibitem{ack} M. Ackermann et al. Science {\bf 339}, 807 (2013).
\bibitem{abdo2} A. Abdo et al. ApJ {\bf{718}}, 348 (2010).
\bibitem{abdo3} A. Abdo et al. Science {\bf{327}}, 1103 (2010).
\bibitem{cas} D. Castro and P. Slane ApJ {\bf{717}}, 372 (2010).
\bibitem{gio} F. Giordano et al. ApJ {\bf{744}} L2 (2012).
\bibitem{ahlers} M. Ahlers, P. Mertsch and S. Sarkar, Phys. Rev. D {\bf 80}, 123017 (2009).
\bibitem{gabici} S. Gabici, F. A. Aharonian and S. Casanova, MNRAS {\bf 396}, 1629 (2009).
\bibitem{ginz} V. L. Ginzburg and S. I. Syrovatskii, 1964, Origin of Cosmic Rays, Pergamon Press, London.
\bibitem{aha1} F. A. Aharonian and A. M. Atoyan, A \& A {\bf 309}, 917 (1996).
\bibitem{kelner} S. R. Kelner, F. A. Aharonian and V. V. Bugayov, Phys. Rev. D {\bf}, 034018 (2006); Erratum-ibid Phys. Rev D {\bf 79}, 039901 (2009).
\bibitem{para} https://sites.google.com/a/ncsu.edu/supernova-remnants.

\end{thebibliography}
\end{document}